\documentclass[12pt]{article}
\usepackage{graphicx}
\usepackage[dvips]{epsfig}

\def\be{\begin{equation}}
\def\ee{\end{equation}}
\def\bea{\begin{eqnarray}}
\def\eea{\end{eqnarray}}

\input{tcilatex}

\begin{document}

\title{Orbifolds, Quantum Cosmology, and Nontrivial Topology\thanks{%
Contribution to the \textit{XXVII National Meeting on the Physics of
Partificles and Fields}, \'{A}guas de Lind\'{o}ia, September 2006.}}
\author{Helio \ V. Fagundes \\
Instituto de F\'{\i}sica Te\'{o}rica, Universidade Estadual Paulista\\
S\~{a}o Paulo, SP 01405-900, Brazil\\
and \and Te\'{o}filo Vargas \\
Instituto de Ci\^{e}ncias Exatas, Universidade Federal de Itajub\'{a}\\
Itajub\'{a}, MG\ 37500-903, Brazil}
\maketitle

\begin{abstract}
In order to include nontrivial spatial topologies in the problem of
quantum creation of a universe, it seems to be necessary to generalize the
sum over compact, smooth 4-manifolds to a sum over finite-volume, compact
\textit{4-orbifolds}. We consider in detail the case of a 4-spherical
orbifold with a cone-point singularity. This allows for the inclusion of a
nontrivial topology in the semiclassical path integral approach to quantum
cosmology, in the context of a Robertson-Walker minisuperspace.
\end{abstract}

\section{Introduction}

The question of whether our universe has a finite or infinite spatial
extension is still an open question, related to that of the global topology
of the universe. The Einstein field equations and the demand for homogeneity
and isotropy deal only with local geometrical properties, leaving the global
topology of the universe undetermined~\cite{lalu}. Thus in the standard
cosmological framework, the universe is described by a
Friedmann-Lamaitre-Robertson-Walker (FLRW) solution, where the spatial
sections are usually assumed to be simply connected. \ These have infinite
volumes in the cases of \ hyperbolic space $H^{3}$ and Euclidean space $%
R^{3} $, and a finite volume in the case of spherical space $S^{3}$. But each of
these geometries can support many topologies with nontrivial
topologies and finite volumes, without altering the dynamics or the
curvature. Therefore there is no reason, either observational or theoretical,
why the universe should not choose a multiply connected topology.

In classical general relativity, a theorem due to Geroch~\cite{ge} and
Tipler~\cite{ti} states that changes of cosmic topology are forbidden, in
the sense that they would imply the appearance of either singularities or
closed timelike curves. Then a natural question appears: where do all
kinds of possible cosmic topologies come from? Quantum cosmology aims at
getting some insight into this problem, by treating the whole universe
quantum mechanically at the Planck era. In fact, it was argued in~\cite{fm}
that the global topology of the present universe would be a relic of its
quantum era, and it should be statistically predictable by quantum
cosmology. Thus, the global topology of cosmic spacetime would not have
changed after the Planck era, and the topology of the present universe
should be the same as that just after the Planck era.

In the pioneering works on the quantum creation of a closed universe, in
both the ``tunneling from nothing''~\cite{tun} and ``no-boundary''~\cite{hh}
proposals, the solution of \ Einstein's equations is a Riemannian 4-sphere $%
S^{4}$ instanton, joined across an equator to a Lorentzian de Sitter space $%
S^{3}$ at its minimal radius. Thus, these and subsequent works on the birth
of a closed universe have been limited to the description of an $S^{4}$
instanton tunneling into a simply connected, Lorentzian de Sitter spacetime $%
R\otimes S^{3}$.

Recently many works have appeared, studying the possibility that the universe
may possess as spatial sections a compact manifold $M$ with nontrivial
topology. $M$ may be represented as a quotient space $R^{3}/\Gamma $, $%
S^{3}/\Gamma $, and $H^{3}/\Gamma $, where $\Gamma $ is a \ (nontrivial)
discrete group of isometries, which is isomorphic to the fundamental group
of $M$. Many methods have been proposed to detect or observationally
constrain the spatial topology, using the prediction of topological images,
catalogs of discrete sources, the search for circles in the sky, and the
cosmic microwave backround radiation.

From this analysis, it is worth considering the quantum creation of the
universe with a nontrivial topology. The flat universe with 3-torus space
topology has been treated in \cite{zel}, while the
hyperbolic case was recently studied by Gibbons, Ratcliffe, and Tschantz 
\cite{gib}. Here we consider the case of a closed universe with $%
S^{3}/\Gamma $ spatial topology. In order to obtain such a universe, it is
necessary to begin with a more general instanton, namely $%
S^{4}/\Gamma $. As shown in \cite{cofa}, the action of $\Gamma $ on $S^{4}$
is not free, so the instanton $S^{4}/\Gamma $ is not a manifold, but rather
an \textit{orbifold } with two cone points, which are the poles of $S^{4}$
and are fixed points \cite{scot} with respect to the nontrivial elements of $%
\Gamma $. Because of this, we generalize Euclidean quantum gravity on
smooth, compact 4-manifolds to Euclidean quantum gravity on finite-volume,
compact 4-orbifolds. Then we obtain the probability of quantum
creation of this universe.

In Sec. 2 we give the general definition of a Riemannian orbifold. In Sec. 3
the Euclidean Einstein-Hilbert action on the orbifold, and the probability
amplitude for the transition from an initial\textit{\ 3-manifold }$\Sigma
_{i}$ to a final\textit{\ 3-manifold}$\ \Sigma _{f}$ are obtained. In Sec. 4
the probability of creation of a universe with the space topology $%
S^{3}/\Gamma $ is calculated, both with the Hartle-Hawking boundary
condition and with the tunneling prescription. And in the last section some
final comments are made.

\section{Orbifold instanton}

An orbifold $O$ under our consideration is a compact, smooth manifold
outside of a finite number of singular points in $O$, and near each of the
latter it is locally homeomorphic to the orbit space of group $\Gamma $. The
notion of orbifold was first introduced by Satake in~\cite{satake}, who used
for it the term \textit{V-manifold}, and was rediscovered by Thurston in~
\cite{thurston}, where the term \textit{orbifold} was coined. Roughly
speaking, an \textit{n-}dimensional manifold is a topological space locally
modeled on Euclidean space ${R}^{n}$, whereas an \textit{n-}orbifold
generalizes this notion by allowing the space to be modeled on quotients on
${R}^{n}$ by finite group action. Similarly for a manifold with boundary vs. an
\textit{orbifold with boundary}, whose formal definition is given below.

\textbf{Definition 1.} Let $X$ be a Hausdorff space. For an open set $U$ in $%
X$, an orbifold coordinate chart over $U$ is a triple ($\tilde{U},\Gamma
,\psi $) such that: $i)$ $\tilde{U}$ is a connected open subset of the
positive half-space ${R}_{+}^{n}=\{(x^{1},x^{2},...,x^{n})\in {R}%
^{n}:x^{n}\geq 0\}$, $ii)$ $\Gamma $ is a finite group of diffeomorphisms
acting on $\tilde{U}$ such that the fixed point set of any $\gamma \in
\Gamma $ which does not act trivially has codimension at least $2$ in $%
\tilde{U}$, and $iii)$ $\psi $: $\tilde{U}\longrightarrow U$ is a continuous
map such that $\forall \gamma \in \Gamma $, $\psi \circ \gamma =\psi $
induces a homeomorphism $\tilde{\psi}:\tilde{U}/\Gamma \longrightarrow U$.

Now suppose that $U$ and $U^{\prime }$ are two open sets in a Hausdorff
space $X$ with $U\subset U^{\prime }$. Let ($\tilde{U},\Gamma ,\psi $) and ($%
\tilde{U}^{\prime },\Gamma ^{\prime },{\psi }^{\prime }$) be charts over $U$
and $U^{\prime }$, respectively.

\textbf{Definition 2.} An injection $\lambda :(\tilde{U},\Gamma ,\psi
)\hookrightarrow (\tilde{U}^{\prime },\Gamma ^{\prime },\psi ^{\prime })$
consists of an open embedding $\lambda :\tilde{U}\hookrightarrow \tilde{U}%
^{\prime }$ such that $\psi =\psi ^{\prime }\circ \lambda ,$ and for any $%
\gamma \in \Gamma $ there exists $\gamma ^{\prime }\in \Gamma ^{\prime }$
for which $\lambda \circ \gamma =\gamma ^{\prime }\circ \lambda $.

An orbifold atlas $\mathcal{U}$ on $X$ is a collection $\{(\tilde{U}%
_{i},\Gamma _{i},\psi _{i})_{i\in \Lambda }\}$, where $\Lambda $ is a
countable set of indices, of compatible orbifold charts with boundary on $X,$
such that $\cup _{i\in \Lambda }\tilde{U}_{i}=X$, and boundary defined by $%
\{\psi _{i}(\partial \tilde{U}_{i}):  (\tilde{U}_{i},\Gamma _{i},\psi _{i})\in \mathcal{U}\}$.

\textbf{Definition 3.} A smooth ${C^{\infty }}$-orbifold with boundary $(X,%
\mathcal{U})$ consists of a Hausdorff space X together with an atlas of
orbifold charts $\mathcal{U}$ satisfying the following conditions: $i)$ For
any pair of charts ($\tilde{U},\Gamma ,\psi $) and ($\tilde{U}^{\prime
},\Gamma ^{\prime },{\psi }^{\prime }$) in $\mathcal{U}$ with $U\subset
U^{\prime }$ there exists an injection $\lambda :(\tilde{U},\Gamma ,\psi
)\hookrightarrow (\tilde{U}^{\prime },\Gamma ^{\prime },{\psi }^{\prime })$.
$ii)$ The open sets $U\subset X$ for which there exists a chart ($\tilde{U}%
,\Gamma ,\psi $) in $\mathcal{U}$ form a basis of open sets in $X$. Given an
orbifold $(X,\mathcal{U})$, we will call the topological space $X$ the 
\textit{underlying space} of the orbifold, and from now on, orbifolds $(X,%
\mathcal{U})$ will be denoted simply by $O$.

Now, let us take a point $x$ in an orbifold $O$ and let ($\tilde{U},\Gamma
,\psi $) be a coordinate chart about $x$. Let $\tilde{x}$ be a point in $%
\tilde{U}$ such that $\psi (\tilde{x})=x$ and let $\Gamma _{\tilde{x}}^{U}$
denote the isotropy group of $\tilde{x}$ under the action of $\Gamma $. As
it is known, the group $\Gamma _{x}^{U}$ depends only on $x$ and not on a
particular choice of $\tilde{x}$ or the chart around $x$. This group $\Gamma
_{x}^{U}$, denoted by $\Gamma _{x}$ is the isotropy group of $x$. Thus, the
singular set $\Im $ of an orbifold consists of those points $x\in O$ whose
isotropy group $\Gamma _{x}$ is nontrivial, and the genuine manifold can be
viewed as an orbifold for which all points have trivial isotropy. We say
that $\Gamma $ acts freely, if for all $x\in M$, $\gamma x=x$ implies $%
\gamma =1$; then the quotient space is another manifold. From this
discussion, $O-\Im $ is an ordinary manifold with boundary.

Thus, a smooth \textit{n-}dimensional orbifold $O$ is a topological space
which locally has the structure of a quotient space of $R^{n}$ by a smooth
finite group action, and its singular set $\Im $ corresponds to the fixed
points of these local actions. For our purpose here, the boundary $\partial
O $ is required to be an $n-1$ compact, smooth manifold without boundary,
which implies that the singular sets are in the interior of $O$. An orbifold
is compact if its underlying topological space is compact.\qquad

The following proposition, due to Thurston, leads to an important
classification of orbifolds into two types:

\textbf{Proposition.} If $M$ is a smooth manifold with boundary and $\Gamma $
a group that acts properly discontinuously on $M$ such that the fixed point
set of each element of $\Gamma $ has codimension greater than or equal to
two and fixes the boundary, then the quotient space $M/\Gamma $ is an
orbifold with boundary. An orbifold is called \textit{good} or \textit{global%
} if it arises as a global quotient of a manifold by a properly
discontinouos group action. Otherwise the orbifold is called \textit{bad}.

Let $O=M/\Gamma $ be a good orbifold covered by $\{(\tilde{U}_{i},\Gamma
_{i},\psi _{i})_{i\in \Lambda }\}$. $O$ is orientable if, for any
overlapping charts $(\tilde{U}_{i},\Gamma _{i},\psi _{i})$ and $(\tilde{U}%
_{j},\Gamma _{j},\psi _{j})$, there exist local manifold coordinate charts $(%
\tilde{x}_{1},\tilde{x}_{2},....,\tilde{x}_{n})$ on $\tilde{U}_{i}$ and $(%
\tilde{y}_{1},\tilde{y}_{2},....,\tilde{y}_{n})$ on $\tilde{U}_{j}$, such
that for any injection $\lambda $ we have $J=\det (\partial \tilde{y}%
_{j}\circ \lambda /\partial \tilde{x}_{i})>0$. Thus an orbifold is
orientable if it has an atlas such that $\tilde{U}_{i}$ can all be oriented
consistently with $\Gamma _{i}$ and the overlap maps are orientation
preserving.

Let us introduce a Riemannian structure on orbifolds. The construction of a
Riemannian metric on a orbifold proceeds, as in the manifold case, with the
metric being locally defined via coordinate charts, and then patched
together through a partition of the unity. The main difference in the
orbifold case is that the structure involved must be invariant under the
local group actions. Let $(X,\mathcal{U})$ be a ${C^{\infty }}$-Riemannian
orbifold with a $\Gamma $-invariant metric. By definition, to give a
Riemannian metric $(,)$ on $(X,\mathcal{U})$ is to give a $\Gamma $%
-invariant Riemannian metric $(,)_{\tilde{U}}$ on each $(\tilde{U}%
_{i},\Gamma _{i},\psi _{i}),$ such that for any injection $\lambda $ of $(%
\tilde{U},\Gamma ,\psi )$ into $(\tilde{U}^{\prime },\Gamma ^{\prime },\psi
^{\prime })$, $(\tilde{X},\tilde{Y})_{\tilde{U}}=\left( \lambda (\tilde{X}%
),\lambda (\tilde{Y})\right) _{\tilde{U}^{\prime }}$, where $\tilde{X}$, $%
\tilde{Y}$ are vector fields on $\tilde{U}$, and $\lambda (\tilde{X})$, $%
\lambda (\tilde{Y})$ are the corresponding vector fields on $\lambda (\tilde{%
U})\subset \tilde{U}{^{\prime }}$. The ${C^{\infty }}$-Riemannian orbifold $%
(X,\mathcal{U})$ has the Levi-Civita connection $\nabla $ defined by a $%
\Gamma $-invariant Riemannian connection ${\nabla }_{\tilde{U}}$ on each $(%
\tilde{U}_{i},\Gamma _{i},\psi _{i}),$ such that for any injection $\lambda $
of $(\tilde{U},\Gamma ,\psi )$ into $(\tilde{U}^{\prime },\Gamma ^{\prime
},\psi ^{\prime })$, $\lambda ({\nabla }_{\tilde{U}\tilde{X}}\tilde{Y})={%
\nabla }_{\tilde{U}^{\prime }\lambda (\tilde{X})}\lambda (\tilde{Y})$. Then
the curvature tensor of $\ {\nabla }$ is defined by 
\begin{equation}
(R_{\tilde{U}})(\tilde{X},\tilde{Y})\tilde{Z}={\nabla }_{\tilde{U}\tilde{X}}{%
\nabla }_{\tilde{U}\tilde{Y}}\tilde{Z}-{\nabla }_{\tilde{U}\tilde{Y}}{\nabla 
}_{\tilde{U}\tilde{X}}\tilde{Z}-{\nabla }_{\tilde{U}[\tilde{X},\tilde{Y}]}%
\tilde{Z}\ ,  \label{curva}
\end{equation}
for vector fields $\tilde{X}$, $\tilde{Y}$, $\tilde{Z}$ on each $\tilde{U}$.
Now, patching the local metrics together using a partition of unity, we
obtain a global Riemannian metric on $O$. A compact orbifold together with a
Riemannian metric is called a \textit{Riemannian orbifold}. From the
curvature tensor the scalar curvature $R[g(x)]$ is obtained as usual, by a
process of contraction.

The next step is to introduce the appropriate notion of integration on an
orbifold. Suppose that $O$ is a compact and orientable Riemannian orbifold.
Let $w$ be an \textit{n-}form on $O$ such that the support of $w$ is
contained in the chart $(\tilde{U},\Gamma ,\psi )$. Then the integral of $w$
on $O$ is defined as follows~\cite{satake}: 
\begin{equation}
\int_{O}w=\frac{1}{|\Gamma |}\int_{\tilde{U}}\tilde{w}\,,
\end{equation}
where $\tilde{w}$ is the pullback of $w$ via the projection $\psi $: $\tilde{%
w}=w\circ \psi $. This definition is independent on the choice of coordinate
chart and the integral of a globally defined differential form is defined
using a partition of unity, as in the manifold case. In general the integration
on a orbifold is defined using the measure $d\mu (g)=\sqrt{g(x)}d^{n}x$
associated with the Riemannian metric $g(x)$ on the orbifold: for an
integrable function $f$, $\int_{O}f\equiv \int_{O}d\mu \lbrack g(x)]f(x).$
Thus we have defined the metric, the curvature, the connection, and the
integration on an$\ n$-dimensional compact, Riemannian, good orbifold $%
O=M/\Gamma $ with boundary $\partial O$. The Einstein-Hilbert functional
(which appears in quantum gravity) on this compact orbifold may be written
as
\begin{equation}
S[g]=\int_{O}d\mu (g)\,R[g]+\int_{\partial O}d\mu (h)\,K[h]\,,  \label{func}
\end{equation}
where $R$ is the scalar curvature of $g$ and $K$ is the trace of the
extrinsic curvature of the boundary hypersurface $(\Sigma ,h)$ in $(O,g)$.
If $\Gamma $ is a discrete subgroup of the isometries of either a spherical,
a hyperbolic, or a Euclidean \textit{n-}manifold, then the quotient spaces $%
S^{n}/\Gamma $, $H^{n}/\Gamma ,$ and $R^{n}/\Gamma $ will be respectively
called a spherical, a hyperbolic and a Euclidean orbifold. The
induced metric and the curvature will have singularities at the fixed points
of the orbifold. However, as observed by Schleich and Witt~\cite{witt}, the
character of the curvature singularity at these points is dimension
dependent; and in dimensions greater than two the integral leaves out only a
set of zero measure, so the Einstein-Hilbert functional is \textit{finite}.

\section{Euclidean quantum gravity and orbifolds}

Following Wheeler's and Hawking's seminal ideas, we formulate a Euclidean
appproach to quantum gravity on orbifolds. Now the basic quantity of
interest is the transition probability from an initial 3-orbifold\textit{\ }%
to a final 3-orbifold, and for $n=4$ the theory of cobordism guarantees that
every compact, oriented 3-orbifold bounds an oriented, compact 4-orbifold.
But we are interested in explaining the origin of our universe, which we
assume to be described by a de Sitter model during the inflationary era and
later by a Friedmann-Robertson-Walker model. Since these models are based on
manifolds the final 3-orbifold $\Sigma _{f}$ is a manifold. The transition
probability amplitude from an initial 3-manifold $\Sigma _{i}$ with a metric 
$h_{i}$ to a final 3-manifold $\Sigma _{f}$ with a metric $h_{f}$ is then
given by 
\begin{equation}
K(\Sigma _{f},h_{f};\Sigma _{i},h_{i},)=\sum_{(O,g)}\int D(g_{\mu \nu })\exp
[-S_{E}(O,g)]\,,  \label{propa}
\end{equation}
where the sum includes any 4-dimensional compact orbifold $O$ with metric $g$;
the integral is over all Lorentz signature $4$-metrics on $O$ which
interpolate between $(\Sigma _{i},h_{i})$ and $(\Sigma _{f},h_{f},)$; and $S_{E}$
is the Euclidean orbifold action with cosmological constant,
\begin{equation}
S_{E}[O_{E},g]=-\frac{1}{16\pi G}\int_{O_{E}}d^{4}x\,g^{1/2}\,(R-2\Lambda )-%
\frac{1}{8\pi G}\int_{\partial O}d^{3}x\,h^{1/2}\,K\,,  \label{aco}
\end{equation}
where $\partial O$ is the union of the disjoint pair of boundaries: $%
\partial O=\Sigma _{i}{\cup }\Sigma _{f}$. By definition, the orbifolds are
4-dimensional, compact, Riemannian, good orbifolds $O=M/\Gamma $ - cf. Sec.
2;  but the important point is that these quotients of $M$ by $\Gamma $ will
induce a nontrivial topology on the boundary $\partial O$. So the transition
probability is from an initial three-orbifold $\Sigma _{i}$ with \textit{%
nontrivial topology} to a final three-manifold $\Sigma _{f}$ also with a
nontrivial topology.

Following the Hartle and Hawking's (HH) proposal~\cite{hh} of
no-boundary boundary condition for the creation of a universe, in which the
initial $\Sigma _{i}$ is absent, Eq. (\ref{propa}) is the wave function for
the metric on\ the 3-manifold boundary $\Sigma =$ $\Sigma _{f}$. Then the
wave function for the universe is defined by a Euclidean sum-over-histories
of the form
\begin{equation}
\Psi (\Sigma ,h_{ij})=\sum_{(O,g)}\int D(g_{\mu \nu })\exp [-S_{E}(O,g)]\,.
\label{psi}
\end{equation}
The path integral (\ref{psi}) has no precise definition, but its qualitative
behaviour can be obtained by using semiclassical techniques in simple
cosmological models, or, equivalently, by restricting the summation to
minisuperspace. The boundary condition in the the semiclassical
approximation to the wave function is of the form
\begin{equation}
\Psi (h_{ij})=N_{0}\sum \,A_{i}\exp (-B_{i})\,,  \label{psi1}
\end{equation}
where $N_{0}$ is a normalization constant and $B_{i}$ are the actions of the
Euclidean classical solutions. The prefactors $A_{i}$ denote fluctuations
about the latter. Thus the wave function for the orbifold instanton in the
semiclassical approximation has the same form as in the case of manifold
instantons. In order to consider the quantum creation of a closed universe
with nontrivial topology, let us outline the main ideas about ${S}%
^{4}/\Gamma $ obtained in~\cite{cofa}. First, let us use the embedding of
the Euclidean four-sphere ${S}^{4}$ with unit radius in 5-dimensional
Euclidean space $R^{5}$: $S^{4}=\{X_{\alpha },\alpha =0,4;X^{\alpha
}X_{\alpha }=1\}$. The action of the discrete, finite group of isometries $%
\Gamma $ on $S^{4}$ is obtained by extending its action on the unit radius
3-sphere $S^{3}$ to all infinite ``parallel'' $S^{3}$ on the four-sphere $%
S^{4}$, that is, for $|X_{0}|\leq 1$, $S_{X_{0}}^{3}=%
\{X_{0},X_{i},i=1,4;X^{i}X_{i}=1-X_{0}^{2}\}$. Thus, the action is already
defined on the equator $S_{0}^{3}$, which is isometric to $S^{3}$. Let $%
(X_{0},X_{i})$ $\in $ ${S}_{X_{0}}^{3}$ and $\gamma \in \Gamma $. If $%
|X_{0}|<1$, then $(0,X_{i}^{^{\prime }}=X_{i}/\sqrt{1-X_{0}^{2}})\in
S_{0}^{3}$, so that $\gamma (0,X_{i}^{^{\prime }})=(0,X_{i}^{^{\prime \prime
}})\in S_{0}^{3}$; we define $\gamma (0,X_{i})\equiv \left(
X_{0},X_{i}^{^{\prime \prime }}\sqrt{1-X_{0}^{2}}\right) \in S_{0}^{3}$. If $%
|X_{0}|=1$, then $\gamma S_{\pm 1}^{3}=S_{\pm 1}^{3}$, which are the poles
of $S^{4}$. Thus the action of $\Gamma $ on $S^{4}$ is not free, and the
quotient space $S^{4}/\Gamma $ is an orbifold (cf. Sec. 2 above) with two
cone-like points corresponding to the poles of the 4-sphere - see Fig. 1.

\begin{figure}[h]
\centering\includegraphics[width=8cm,height=10cm]{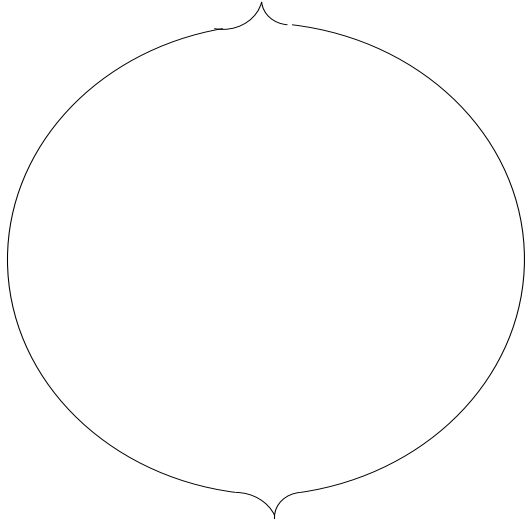}
\caption{The orbifold instanton ${S}^{4}/\Gamma $ with the poles of the 4-sphere
as cone points.}
\end{figure}

In the quantum creation of a universe, in both the tunneling from nothing
and the no-boundary proposals, only the lower half $(X_{0}\leq 0)$ of the
orbifold instanton takes part in the solution. The full spacetime solution
is $M=O\,{\cup }_{\Sigma }\,M_{L}$, where $O$ and $M_{L}=R\otimes
S^{3}/\Gamma $ are attached smoothly by $\Sigma =S_{0}^{3}/\Gamma =\partial
O\ $(see Fig. 2). Thus Gibbons's condition~\cite{gib} is satisfied: $O$ is a
compact orbifold with $\Sigma $ as sole boundary; $\Sigma $ is a possible
Cauchy surface for $M_{L}$, and has a vanishing second fundamental form with
respect to both $O$ and $M_{L}$\thinspace - \thinspace this is true for the $%
S^{3}$ covering, and the action of $\ \Gamma $ does not interfere with the
local metrics.

\begin{figure}
\centering\includegraphics[width=8cm,height=12cm]{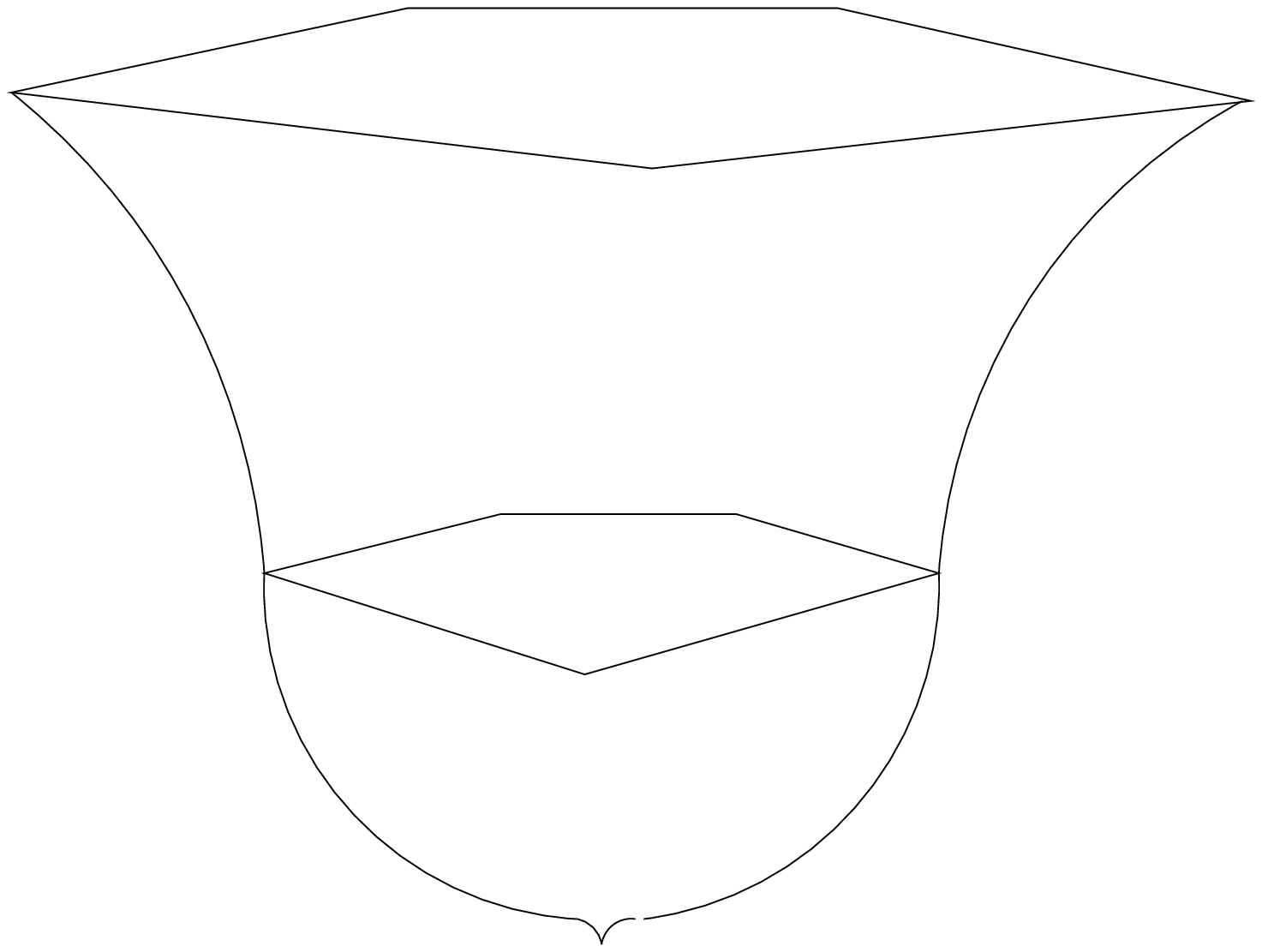}
\caption{Global estructure of the orbifold instanton and its continuation.
The orbifold instanton ${S}^{4}/\Gamma $ is joined across an equator to a
Lorentzian de Sitter space with nontrivial topology ${S}^{3}/\Gamma $ at its
radius of maximum contraction.}
\end{figure}

\section{Calculation of the wave function}

In this section we will consider the probability of creation of a universe
with the space topology $S^{3}/\Gamma $ in both the Hartle-Hawking
no-boundary condition and the tunneling prescription. In order to calculate
the minisuperspace path integral the semiclassical approximation given by
Eq. (\ref{psi1}) is used.

The line element of a Euclidean Robertson-Walker (RW) universe is 
\begin{equation}
ds^{2}=N(\tau )^{2}d\tau ^{2}+a^{2}(\tau )\left[ \frac{dr^{2}}{1-kr^{2}}%
+r^{2}d\Omega _{2}^{2}\right] .
\end{equation}
The (comoving) volume of the space section $\Sigma =S^{3}/\Gamma $ is given
by 
\begin{equation}
V(S^{3}/\Gamma )=\frac{2\pi ^{2}}{|\Gamma |},
\end{equation}
where $|\Gamma |$ is the order of $\Gamma $, so the Euclidean
Einstein-Hilbert action (\ref{aco}) is 
\begin{equation}
S_{E}=\frac{3V(S^{3}/\Gamma )}{8\pi G}\int\limits_{{\tau }_{0}}^{{\tau }%
_{1}}\,d\tau \left[ -\frac{\dot{a}^{2}a}{N}-Na+\frac{Na^{3}\Lambda }{3}%
\right] .  \label{aco1}
\end{equation}
Thus an information about the topology of the 4-orbifold is encoded in the
action.

The field equation is
\begin{equation}
\left( \frac{\dot{a}}{N}\right) ^{2}-1+\frac{\Lambda }{3}a^{2}=0\,,
\end{equation}
whose solution
\[
a(\tau )=\frac{1}{H}\cos (HN\tau )\ ,
\]
with $H=\sqrt{\Lambda /3}$ , satisfies the boundary conditions $a(\tau )=0$
at the south pole of $O_{E}$ and $a(\tau )=a_{0}=$ radius both of the
Lorentzian de Sitter space with topology ${S}^{3}/\Gamma $ and of $O_{E}$.

Using this boundary condition, the final Euclidean action is
\begin{equation}
S_{E}=-\frac{V(S^{3}/\Gamma )}{6\pi GH^{2}}\left[ 1-\left(
1-H^{2}a_{0}^{2}\right) ^{3/2}\right] \,.
\end{equation}
The semiclassical approximation of the HH wave function (\ref{psi1}) for $%
a<H^{-1}$ is therefore
\begin{equation}
\Psi (S^{3}/\Gamma )=N_{0}\exp \left( \frac{\pi }{3|\Gamma |GH^{2}}%
\left[ 1-\left( 1-H^{2}a^{2}\right) ^{3/2}\right] \right) \ ,
\end{equation}
and the unnormalized probability of creation of this multiply connected
universe is
\begin{equation}
|\Psi _{HH}|^{2}=N_{1}\exp \left( \frac{2\pi }{3|\Gamma |GH^{2}}\left[
1-\left( 1-H^{2}a^{2}\right) ^{3/2}\right] \right) .  \label{whh}
\end{equation}

On the other hand, Vilenkin~\cite{vile} observed that the full
Wheeler-DeWitt equation is invariant under the transformation 
\begin{equation}
h_{ij}\rightarrow \exp {(i\pi )}h_{ij}\,,\ \quad V(\phi )\rightarrow \exp {%
(-i\pi )}V(\phi )\ .
\end{equation}
This means that there is a relation between the tunneling and the HH wave
funtions: $\Psi _{HH}$ and $\Psi _{T}$ are related by an analytical
continuation.

Using these relations for our case, we obtain the tunneling wave function
for $a<H^{-1}$
\begin{equation}
\Psi _{T}(S^{3}/\Gamma )=N_{0}\exp \left( -\frac{\pi }{3|\Gamma |GH^{2}}%
\left[ 1-\left( 1-H^{2}a^{2}\right) ^{3/2}\right] \right) \,,
\end{equation}
and the unnormalized probability of creation of a universe in the tunneling
approach, for $a<H^{-1}$ is 
\begin{equation}
|\Psi _{T}|^{2}=N{}_{1}\exp \left( -\frac{2\pi }{3|\Gamma |GH^{2}}\left[
1-\left( 1-H^{2}a^{2}\right) ^{3/2}\right] \right) \ .  \label{tu}
\end{equation}
Thus we obtain the probability of quantum creation of a universe with an $%
S^{3}/\Gamma $ spatial topology, in both the Hartle-Hawking (HH) boundary
condition and Vilenkin's tunneling prescription.

\section{Final remarks}

We have considered the generalization of Euclidean quantum gravity on
compact, smooth 4-manifolds to a Euclidean quantum gravity on finite-volume,
compact 4-orbifolds, and as an application we obtained the probability of
quantum creation of a closed universe with an $S^{3}/\Gamma $ spatial
topology. Using this Euclidean functional integral prescription as applied
to quantum cosmology, we calculated the wave function of such a universe
with a positive cosmological constant and without matter. The minisuperspace
path integral is calculated in the semiclassical approximation and thus we
obtained the probability of quantum creation of a closed universe with
nontrivial topology, and comparing Eqs. (\ref{whh}) and (\ref{tu}), it
appears that in HH approach the probability of creation is \textit{maximum}
for minimum order of $\Gamma $, while in the tunneling approach it \textit{%
increases} with $|\Gamma |$.

Further generalizations, including a cosmological constant and a scalar field of
matter, both in the above spherical case and in that of a compact, hyperbolic orbifold
instanton $H^{4}/\Gamma $ going over to a compact, hyperbolic FLRW universe, are under
study.

\section*{Acknowledgments}

One of us (T. V.) would like to thank FAPESP and FAPEMIG for financial
support.

\end{document}